\documentstyle[11pt]{article}
\titlepage

\title{Determination of the Metric from the Connection}
\author{  Richard Atkins \\
        richard.atkins@twu.ca \\
		Department of Mathematics\\ 
		Trinity Western University \\
		7600 Glover Road \\
		Langley, BC, V2Y 1Y1 Canada}
\date{}
\newtheorem{fact}{Fact}

\newtheorem{lemma}[fact]{Lemma}

\newtheorem{theorem}[fact]{Theorem}
\newtheorem{corollary}[fact]{Corollary}
\begin{document}
\maketitle
\begin{abstract}
As is well known, a metric on a manifold determines a unique symmetric connection
for which the metric is parallel: the Levi-Civita connection. In this paper we
investigate the inverse problem: to what extent is the metric of a Riemannian manifold  
determined by its Levi-Civita connection? It is shown that for a generic Levi-Civita
connection of a metric $h$ there exists a set of positive semi-definite tensor 
fields $h_{a}$ such that the parallel metrics are the positive-definite 
linear combinations of the $h_{a}$. Moreover, the set of all parallel metrics 
may be constructed by a soley algebraic procedure.
\end{abstract}

\newpage

\section{Introduction}

One of the fundamental principles of differential geometry is that a Riemannian manifold 
$(M,h)$ uniquely determines a connection: the Levi-Civita connection of the metric. 
This paper examines how much information on the metric $h$ is retained by the associated 
Levi-Civita connection. Specifically, we inquire to what extent the metric of a Riemannian 
manifold is determined by the Levi-Civita connection and seek a method for constructing
the general form of a parallel metric on the manifold. 

We begin, in the following section, by introducing a genericity condition on the set of 
connections, defined in terms of the Riemann curvature tensor. This shall enable us 
to investigate the structure of the space of local parallel metrics; that is, metrics defined 
on an open set of the manifold, which are parallel with respect to the given Levi-Civita 
connection $\nabla$. This, in turn, shall lead to a decomposition of the tangent space 
of the manifold into a direct sum of orthogonal subbundles. 
The original metric $h$, restricted to the subbundles of the decomposition, defines a set of
positive semi-definite tensor fields $h_{a}$. It is proved that a metric is parallel with 
respect to $\nabla$ if and only if it is a positive-definite linear combination of the 
$h_{a}$.  Furthermore, the complete set of parallel metrics may be obtained 
by a purely algebraic procedure; integration of differential equations is not required.
Lastly, we provide an example that illustrates the method described herein.

In \cite{aa}, it is determined when an arbitrary analytic symmetric connection is a 
Levi-Civita connection for some metric. 
The problem of finding the metric from the Ricci curvature has been studied by 
DeTurk \cite {bb}, \cite{cc}, \cite{dd} and \cite{ee}, and from 
the curvature in general relativity by Hall \cite{hh} and Hall and McIntosh \cite{ii}.
 
\section{Parallel metrics for generic Levi-Civita connections}

In what follows, $M$ shall denote a connected manifold of dimension $n$. This is not
essential but it shall lead to concepts that are more familiar. Moreover, 
the scope of the paper is not thereby limited since the results below may be applied to 
the separate components of a general manifold.

It will also be convenient to represent a metric in terms of contravariant indices: 
$<,> = g = \sum_{i,j=1}^{n}g^{ij}X_{i}\otimes X_{j}$ where $(X_{1},...,X_{n})$
is a local frame. This shall be the convention throughout.

Define a connection $\nabla$ on $M$  to be {\it generic} at
a point $m \in M$ if there exist tangent vectors $\xi_{1},\xi_{2} \in T_{m}M$
such that the linear transformation
\[ R(\xi_{1},\xi_{2}): T_{m}M \rightarrow T_{m}M \]
has $n$ distinct (complex) eigenvalues, where $R$ denotes the Riemann curvature tensor
of $\nabla$. A connection is {\it generic} if 
it is generic at every point $m \in M$. 
Henceforth $\nabla$ shall denote a generic symmetric connection on $M$ with parallel 
metric $h$.

Consider a connected open set $U$ of $M$ for which there exist vector fields 
$\xi_{1},\xi_{2}$ on $U$ such that 
\[ R(\xi_{1}(u),\xi_{2}(u)):T_{u}M \rightarrow T_{u}M \]
has $n$ distinct eigenvalues $\lambda_{i} = \lambda_{i}(u)$, for each $u\in U$, 
and an associated frame of (complex) eigenvector fields $(Z_{1},...,Z_{n})$ defined on $U$:
\[  R(\xi_{1},\xi_{2})(Z_{i}) = \lambda_{i}Z_{i} \]
Let ${\cal \widehat{U}}$ denote the set of all such connected open sets of $M$.
Since $\nabla$ is generic, ${\cal \widehat{U}}$ is an open covering of $M$. 
If $V \in \widehat{\cal U}$ then any connected open subset $W\subseteq V$ 
is also in $\widehat{\cal U}$.

\begin{lemma} \label{lemma:eigen} \hspace{1in}  \\
(i) The frame of eigenvector fields $(Z_{1},...,Z_{n})$  of $R(\xi_{1},\xi_{2})$,
after a possible reordering,  has the form
 \[  X_{1}+iX_{2}, X_{1}-iX_{2},
...,X_{2m-1}+iX_{2m}, X_{2m-1}-iX_{2m},(X_{n})  \] 
where the $X_{i}$ are vector fields on $U$ and $X_{n}$, 
the eigenvector field corresponding to the zero eigenvalue, is included if $n$ is odd. 
Furthermore,\\
(ii) $X_{1},...,X_{2m},(X_{n})$ is an orthogonal frame on $U$ for any metric $g$  on $U$,
parallel with respect to $\nabla$. That is, $g$ is expressible in the form 
$g = \sum_{i=1}^{n} g_{i}X_{i}\otimes X_{i}$ for some functions $g_{i}:U \rightarrow \Re^{+}$.
\end{lemma} 
{\bf Proof:}\\
(i) Since $\nabla$ is the Levi-Civita connection of the metric $h$, 
there exists an orthonormal basis of $T_{u}M$ at each 
$u\in U$, with respect to which $R(\xi_{1},\xi_{2})$ is represented as a 
skew-symmetric matrix. Hence the eigenvalues $\lambda_{i} = \lambda_{i}(u)$ 
are purely imaginary with associated eigenvector fields $Z_{2k-1}=X_{2k-1}+iX_{2k}$
for $\lambda_{2k-1}$ and $Z_{2k}= \overline{Z}_{2k-1} =X_{2k-1}-iX_{2k}$ for 
$\lambda_{2k}=-\lambda_{2k-1}$, {$1\leq k\leq m$,
except for $\lambda_{n}=0$ when $n$ is odd, with associated eigenvector field $Z_{n}=X_{n}$. \\
(ii) Let $g=\sum_{i,j=1}^{n} g^{ij}Z_{i} \otimes Z_{j}$ be a metric parallel on $U$;
thus $R(\xi_{1},\xi_{2})(g)=0$. The explicit form of $g$ gives 
\begin{eqnarray*}
R(\xi_{1},\xi_{2})(g) & = & [ \nabla_{\xi_{1}}, \nabla_{\xi_{2}} ](g) - 
                           \nabla_{[ \xi_{1}, \xi_{2} ]}(g) \\
       & = & \sum_{i,j=1}^{n} g^{ij}(R(\xi_{1},\xi_{2})(Z_{i})\otimes Z_{j}+
                                     Z_{i}\otimes R(\xi_{1},\xi_{2})(Z_{j}))  \\
       & = & \sum_{i,j=1}^{n} g^{ij}(\lambda_{i}Z_{i}\otimes Z_{j}+
                                     Z_{i}\otimes \lambda_{j}Z_{j}) \\
       & = & \sum_{i,j=1}^{n} g^{ij}(\lambda_{i}+ \lambda_{j})Z_{i}\otimes Z_{j}
\end{eqnarray*}                           
Therefore $g^{ij}(\lambda_{i}+\lambda_{j})=0$ for all $1 \leq i,j \leq n$.
The eigenvalues $\lambda_{i}$ are distinct, by hypothesis, and 
$\lambda_{2k}=-\lambda_{2k-1}$, {$1\leq k\leq m$,
except for $\lambda_{n}=0$ when $n$ is odd. Hence $g^{ij}=0$, unless  
$(i, j) = (2k-1,2k)$ or $(i, j) = (2k,2k-1)$ for some $k \in \{1,...,m\}$, or
$i=j=n$ when $n$ is odd.
It follows that $g^{ij}$ is block diagonal with $2\times 2$ blocks down the 
main diagonal and with a single $1\times 1$ block for odd $n$. In the
$(X_{1},...,X_{n})$ frame the $k^{th}$ $2\times 2$ block has the form
\begin{eqnarray*}
\sum_{i,j=2k-1}^{2k}g^{ij}Z_{i}\otimes Z_{j} & = & 
      g^{2k-1,2k}Z_{2k-1} \otimes Z_{2k} + g^{2k,2k-1}Z_{2k} \otimes Z_{2k-1} \\
   & = &    g^{2k-1,2k}(X_{2k-1}+iX_{2k})\otimes (X_{2k-1}-iX_{2k})+ \\
   & &       g^{2k,2k-1}  (X_{2k-1}-iX_{2k})\otimes (X_{2k-1}+iX_{2k}) \\
   & = &    g_{2k-1}X_{2k-1}\otimes X_{2k-1}+ g_{2k}X_{2k}\otimes X_{2k}
\end{eqnarray*}
where $g_{2k-1}=g_{2k} := g^{2k-1,2k} + g^{2k,2k-1}$. Defining 
$g_{n}:=g^{nn}$ for odd $n$, 
\[ g = \sum_{i=1}^{n} g_{i}X_{i}\otimes X_{i} \]
{\bf q.e.d.}\\

Let $\theta=\theta^{i}_{j}$ denote the 
connection form of $\nabla$ in the $(X_{1},...,X_{n})$ frame:
\[ \nabla X_{j} = \sum_{i=1}^{n} X_{i}\otimes \theta^{i}_{j} \]
The following lemma indicates the amount of variation allowed
among parallel metrics. 

\begin{lemma} \label{lemma:parallel}
Let $<,>^{1} = \sum_{i=1}^{n} g_{i}X_{i}\otimes X_{i}$ and 
$<,>^{2} = \sum_{i=1}^{n} k_{i}X_{i}\otimes X_{i}$ be two arbitrary metrics on $U$, parallel
with respect to $\nabla$. Then there exist constants $c_{i} \in {\Re}^{+}$ such that
$g_{i} = c_{i}k_{i}$, for all $1\leq i \leq n$. 
\end{lemma}
{\bf Proof:} \\
Since $<,>^{1}$ is parallel,
\begin{eqnarray*}
0 & = & \nabla \hspace{0.05in} (\sum_{i=1}^{n} g_{i}X_{i} \otimes X_{i}) \\
  & = & \sum_{i,j=1}^{n} X_{i}\otimes X_{j} \otimes (\delta_{ij}dg_{i} + g_{j}\theta^{i}_{j}
      +g_{i}\theta^{j}_{i})
\end{eqnarray*}
When $i=j$ this gives, $\theta^{i}_{i} = -\frac{1}{2}dlog g_{i}$.
Similarly, $\theta^{i}_{i} = -\frac{1}{2}dlog k_{i}$. Therefore
\[  dlog g_{i} = dlog k_{i} \]
and since $U$ is connected, $g_{i} = c_{i}k_{i}$ for some $c_{i} \in {\Re}^{+}$.
\\
{\bf q.e.d.}\\

Next, we shall seek to make this observation stronger. 
In light of Lemma \ref{lemma:eigen}, the metric $h$ can be written locally as
\[ h|_{U} = \sum_{i=1}^{n}\rho_{i}X_{i}\otimes X_{i} \]
for some functions $\rho_{i} : U \rightarrow \Re^{+}$. Define the orthonormal basis of
local vector fields $(Y_{1},...,Y_{n})$ by  $Y_{i} := \sqrt{ \rho_{i}}X_{i}$, for 
$1\leq i \leq n$. Then  $h|_{U}=Y_{1}\otimes Y_{1}+\cdots +Y_{n}\otimes Y_{n}$.
Let $\omega = \omega^{i}_{j}$ denote the connection form of $\nabla$ with respect to 
the frame $(Y_{1},...,Y_{n})$:
\[ \nabla Y_{j} = \sum_{i=1}^{n} Y_{i}\otimes \omega^{i}_{j}  \]

\begin{lemma} \label{lemma:iff}
Consider $g=\sum_{i=1}^{n}c_{i}Y_{i}\otimes Y_{i}$, where $c_{i} \in \Re$.
$\nabla g = 0$ if and only if $(c_{i}-c_{j})\omega^{i}_{j}(u) = 0 $ for all
$u\in U$ and  $1\leq i,j \leq n$.
\end{lemma}
{\bf Proof:}\\ Taking the covariant derivative of $g$ gives,
\[ \nabla g = \nabla \hspace{0.05in} (\sum_{i=1}^{n} c_{i}Y_{i} \otimes Y_{i}) 
   =  \sum_{i,j=1}^{n} Y_{i}\otimes Y_{j} \otimes (c_{j}\omega^{i}_{j}
      +c_{i}\omega^{j}_{i}) \]
Therefore $\nabla g = 0$ if and only if $c_{j}\omega^{i}_{j}+c_{i}\omega^{j}_{i} = 0$ for all
$1\leq i,j \leq n$. Since $\omega^{i}_{j} = -\omega^{j}_{i}$, this holds if and only if 
$(c_{i}-c_{j})\omega^{i}_{j} = 0 $ for all $1\leq i,j \leq n$.
\\
{\bf q.e.d.} \\

Define $r(U)$ to be the  equivalence 
relation on $\{1,...,n\}$ generated by the relations 
$\{ (i,j) \hspace{.1in} | \hspace{.1in} \omega^{i}_{j}(u) \neq 0$ for some $u\in U \}. $
Thus  $ir(U)j$ for $i\neq j$ 
if and only if there exists a sequence $i = i_{1},...,i_{k} = j$
in $\{1,...,n\}$ and $u_{i_{1}},...,u_{i_{k-1}}\in U$ such that 
$\omega^{i_{l}}_{i_{l+1}}(u_{i_{l}}) \neq 0 $ for all
$1\leq l \leq k -1$. $r(U)$  partitions   $\{1,...,n\}$ into
$\beta(U)$ disjoint subsets $P_{1}(U),...,P_{\beta(U)}(U)$. 

Define tensor fields $h_{i}(U)$  on $U$ by
\[ h_{i}(U) := \sum_{j\in P_{i}(U)}Y_{j}\otimes Y_{j}, \]
for $1\leq i \leq \beta(U)$. 
Note that 
\[ h|_{U}=\sum_{i=1}^{\beta(U)}h_{i}(U) \]

\begin{lemma} \label{lemma:a}
If \[ g = \sum_{i=1}^{\beta(U)} a_{i}h_{i}(U)\]
for constants $a_{i} \in \Re$ then $g$ is parallel with respect to $\nabla$. 
\end{lemma}
{\bf Proof:}\\
Suppose that 
$g = \sum_{i=1}^{n}c_{i}Y_{i}\otimes Y_{i} = \sum_{k=1}^{\beta(U)} a_{k}h_{k}(U) $
for constants $c_{i}, a_{k} \in \Re$. If $ir(U)j$ then $i$ and $j$ belong
to the same equivalence class $P_{k}(U)$, say. Hence $c_{i}=c_{j}=a_{k}$. On the other 
hand, if $i$ and $j$ are not $r(U)$-related then $\omega^{i}_{j}(u)=0$ for all 
$u\in U$. In either case, $(c_{i}-c_{j})\omega^{i}_{j}(u) = 0 $ for all
$u\in U$. Therefore by Lemma \ref{lemma:iff}, $g$ is parallel with respect to $\nabla$.\\
{\bf q.e.d.} 

\begin{lemma} \label{lemma:linear}
If $g$ is a metric on $U$ parallel with respect to $\nabla$ then 
\[ g = \sum_{i=1}^{\beta(U)} a_{i}h_{i}(U)\]
for some constants $a_{i} \in \Re^{+}$. 
\end{lemma}
{\bf Proof:}\\
Let $g$ be a metric on $U$ parallel with respect to $\nabla$. 
By Lemma \ref{lemma:eigen}, $g$ may be written $g = \sum_{i=1}^{n} g_{i}X_{i}\otimes X_{i}$
for some functions $g_{i}:U\rightarrow \Re^{+}$.
Then Lemma \ref{lemma:parallel} and the fact that
$h|_{U} = \sum_{i=1}^{n}\rho_{i}X_{i}\otimes X_{i}$ imply that $g$ is of the form
\[ g = \sum_{i=1}^{n}c_{i}\rho_{i}X_{i}\otimes X_{i} =
     \sum_{i=1}^{n}c_{i}Y_{i}\otimes Y_{i}\]
for some constants $c_{i}\in {\Re}^{+}$. By Lemma \ref{lemma:iff},
$(c_{i}-c_{j})\omega^{i}_{j}(u) = 0 $ for all $u\in U$ and  $1\leq i,j
\leq n$. Hence $c_{i}=c_{j}$ whenever $ir(U)j$. This means that $g$ may be expressed as
 $g = \sum_{i=1}^{\beta(U)} a_{i}h_{i}(U)$ for constants $a_{i} \in \Re^{+}$. \\ 
{\bf q.e.d.} \\

Up to this point we have explored the structure of the set of parallel metrics
on a single open set. We shall now investigate how these structures relate
to each other on intersecting sets. The definition of $h_{i}(U)$ described above
depends upon (1) the choice of vector fields $\xi_{1}$ and $\xi_{2}$ on $U$ enforcing the
genericity condition, (2) the ordering of the associated orthonormal frame 
$(Y_{1},...,Y_{n})$, and (3) the ordering of the blocks, $P_{1}(U),...,P_{\beta(U)}(U)$, 
of the associated partition. For each $U$ in $\widehat{\cal U}$
we shall require that such choices have been made and define the tensor fields
$h_{i}(U)$, $1\leq i\leq \beta(U)$, accordingly. 
 
\begin{lemma} \label{lemma:sum}
Let $W\subseteq U$ be sets in $\widehat{\cal U}$. 
Then there exist constants $c_{iq} \in \Re$ such that
\[ h_{i}(U)|_{W} = \sum_{q=1}^{\beta(W)} c_{iq}h_{q}(W) \]
for $1\leq i \leq \beta(U)$.
\end{lemma}
{\bf Proof:}\\
By Lemma \ref{lemma:linear}, 
\[ \sum_{j=1}^{\beta(U)} h_{j}(U)|_{W} = \sum_{q=1}^{\beta(W)}a_{q} h_{q}(W) \]
for some $a_{q} \in \Re^{+}$.
By Lemmas \ref{lemma:a} and \ref{lemma:linear},
\[ h_{i}(U)|_{W} +\sum_{j=1}^{\beta(U)} h_{j}(U)|_{W} = \sum_{q=1}^{\beta(W)}b_{iq} h_{q}(W) \]
for some $b_{iq} \in \Re^{+}$. Subtracting gives
\[ h_{i}(U)|_{W} = \sum_{q=1}^{\beta(W)} c_{iq}h_{q}(W) \]
where $c_{iq} := b_{iq}-a_{q} \in \Re$. \\
{\bf q.e.d.} \\

Let $\bar{Q}_{i}(U)$ be the subbundle of $TU$ spanned by the vector fields 
$\{Y_{j}: j\in P_{i}(U)\}$, for each $i \in 1,...,\beta(U)$. We then have a  
decomposition of the tangent space $TU = \bar{Q}_{1}(U)\oplus \cdots \oplus 
\bar{Q}_{\beta(U)}(U)$. Denote the dual subbundles by $\bar{Q}^{*}_{i}(U)$. Then
$T^{*}U = \bar{Q}^{*}_{1}(U)\oplus \cdots \oplus \bar{Q}^{*}_{\beta(U)}(U)$.
Observe that $h_{i}(U)$ is a section of $\bar{Q}_{i}(U)\otimes
\bar{Q}_{i}(U)$, defining a positive semi-definite  bilinear form
$h_{i}(U):T^{*}_{u}U\times T^{*}_{u}U \rightarrow \Re$ for each $u\in U$.
Set 
\[ h_{i}(U)^{\perp}:= \bigcup_{u\in U} \{\sigma \in T^{*}_{u}U: h_{i}(U)(\sigma,\tau)=0,
\hspace{0.1in} \forall \tau\in T^{*}_{u}U \} \]
Then
\[  h_{i}(U)^{\perp} = \bigoplus_{j\neq i}\bar{Q}^{*}_{j}(U) \]

\begin{lemma} \label{lemma:subset}
Let $W\subseteq U$ be sets in $\widehat{\cal U}$. Then there exists a partition
\[ \Gamma(1),...,\Gamma(\beta(U)) \] of the set of integers $\{1,...,\beta(W) \}$ such that
\[ \bar{Q}_{i}(U)|_{W} = \bigoplus_{q\in \Gamma(i)} \bar{Q}_{q}(W) \]
and
\[ h_{i}(U)|_{W} = \sum_{q\in \Gamma(i)} h_{q}(W) \]
for $1\leq i \leq \beta(U)$.
\end{lemma}
{\bf Proof:}\\
By Lemma \ref{lemma:sum}, there exist constants $c_{iq} \in \Re$ such that
\[ h_{i}(U)|_{W} = \sum_{q=1}^{\beta(W)} c_{iq}h_{q}(W) \]
for $1\leq i \leq \beta(U)$. Define subsets $\gamma(i)$ and $\Gamma(i)$ of 
$\{1,...,\beta(W) \}$ by
\[ \gamma(i) := \{ q : c_{iq} \neq 0 \} \]
and
\begin{eqnarray*} \Gamma(i) & := & \{ 1,..., \beta(W) \} -  \bigcup_{j \neq i} \gamma (j) \\
                            &  = &     \{ q: c_{jq} = 0, \hspace{0.1in} \forall j\neq i \} 
\end{eqnarray*}
for $1\leq i \leq \beta(U)$. Now,
\begin{eqnarray*}
  \bar{Q}^{*}_{i}(U)|_{W} & = & 
        \hspace{0.08in} \bigcap_{j\neq i} \hspace{0.1in} (h_{j}(U)|_{W})^{\perp} \\
  & = & \hspace{0.08in} \bigcap_{j\neq i} \hspace{0.1in} 
       (\sum_{q=1}^{\beta(W)} c_{jq}h_{q}(W))^{\perp} \\
  & = & \hspace{0.08in}\bigcap_{j\neq i} \hspace{0.1in} 
        (\sum_{q\in\gamma(j)} h_{q}(W))^{\perp} \\
  & = & \hspace{0.08in}\bigcap_{j\neq i} \hspace{0.1in} 
        \bigoplus_{q\notin\gamma(j)} \bar{Q}^{*}_{q}(W) \\
  & = & \bigoplus_{q\in \Gamma(i)}  \bar{Q}^{*}_{q}(W)
\end{eqnarray*} 
Hence,
\[ \bigoplus_{q=1}^{\beta(W)} \bar{Q}^{*}_{q}(W) = T^{*}W =
   \bigoplus_{i=1}^{\beta(U)} \bar{Q}^{*}_{i}(U)|_{W} =
   \bigoplus_{i=1}^{\beta(U)} \hspace{0.1in} \bigoplus_{q\in\Gamma(i)} \bar{Q}^{*}_{q}(W)   \]
It follows that $\Gamma(1),...,\Gamma(\beta(U))$ is a partition of $\{1,...,\beta(W)\}$
and 
\[ \bar{Q}_{i}(U)|_{W} = \bigoplus_{q\in \Gamma(i)} \bar{Q}_{q}(W) \]
for $1\leq i \leq \beta(U)$.

Furthermore,
\begin{eqnarray*}
h_{i}(U)|_{W} & = & \hspace{0.06in} \sum_{q=1}^{\beta(W)} c_{iq}h_{q}(W) \\
              & = & \hspace{0.06in} \sum_{j=1}^{\beta(U)} \hspace{0.1in} 
                    \sum_{q\in \Gamma(j)} c_{iq}h_{q}(W) \\
              & = & \sum_{q\in \Gamma(i)} c_{iq}h_{q}(W)
\end{eqnarray*}
since $c_{iq} = 0$ for $q\in \Gamma(j)$ when $j\neq i$. Consequently,
\[ \sum_{q=1}^{\beta(W)}h_{q}(W) = h|_{W} =
   \sum_{i=1}^{\beta(U)}h_{i}(U)|_{W} =
   \sum_{i=1}^{\beta(U)} \hspace{0.1in} \sum_{q\in \Gamma(i)} c_{iq}h_{q}(W) \]
Therefore $c_{iq}=1$ for all $q\in \Gamma(i)$ and so
\[ h_{i}(U)|_{W} = \sum_{q\in \Gamma(i)} h_{q}(W) \] \\
{\bf q.e.d.} \\

We now turn to the problem of how to piece together these local parallel metrics into 
a global one. This is done by means of an appropriately defined equivalence relation.
Let ${\cal U}$ be any subset of $\widehat{\cal U}$ that covers $M$ and has the property that   
the intersection of pairs of sets in ${\cal U}$ is connected. For instance,
we can choose ${\cal U}$ to be a good refinement of $\widehat{\cal U}$.
Let $I:=\{(i,U):U \in {\cal U}$ and $1\leq i \leq \beta(U)\}$ and define 
$\sim$ to be the equivalence relation on $I$ generated by the relations
\[ (i,U) \sim (j,V) \hspace{0.2in} if \hspace{0,2in} U \cap V \neq \emptyset 
\hspace{0.2in} and \hspace{0.2in} \bar{Q}_{i}(U) \cap \bar{Q}_{j}(V) \neq 0  \]
where "$0$" means the zero distribution on $U\cap V$. 
These relations are required in order to join the
$\bar{Q}_{i}$ distributions together in a smooth way. The equivalence partitions $I$ into
$A$ blocks denoted $I_{1},...,I_{A}$. 

\begin{lemma} \label{lemma:partition}
For each $a\in \{1,...,A\}$ and $U\in {\cal U}$ the set $\{i:(i,U) \in I_{a}\}$
is non-empty.
\end{lemma}
{\bf Proof:}\\
Let $C(U)$ denote the subset of ${\cal U}$ consisting of all sets $U'$ for which
there exists a sequence of sets $U'=U_{1},U_{2},...,U_{k}=U$ in ${\cal U}$ such that
$U_{l}\cap U_{l+1} \neq \emptyset$ for all $1\leq l\leq k-1$. 
We claim that $C(U) = {\cal U}$. First observe that if $V \in {\cal U}$ and
$V \cap U' \neq \emptyset$ for some $U' \in C(U)$ then $V\in C(U)$ also.
Let $S$ denote the union of the sets in $C(U)$. Suppose that $S$ is not equal to $M$.
Since $M$ is connected the boundary $\partial S$, of $S$ is non-empty. 
${\cal U}$ covers $M$, so there exists an open set 
$V\in {\cal U}$ such that $V\cap \partial S \neq \emptyset$.  
This means that $V \cap S \neq \emptyset$ and
$V\cap S^{c} \neq \emptyset$, where $S^{c}$ denotes the compliment of $S$ in $M$. 
Hence $V \cap U' \neq \emptyset$ for some $U' \in C(U)$ and $V \notin C(U)$,
which is a contradiction. Therefore $S=M$. Let $W \in {\cal U}$. Then $W\cap U' \neq
\emptyset $ for some $U' \in C(U)$ and so $W\in C(U)$. This demonstrates the claim.
   
Let $(i_{1},U_{1})$ be any representative of $I_{a}$. 
We have shown that there exists a sequence of sets
$U_{2},...,U_{k}=U$ in ${\cal U}$ such that $U_{l}\cap U_{l+1} \neq \emptyset$
for all $1\leq l\leq k-1$. For each $l \in \{1,...,k-1\}$ and $p \in \{ 1,...,\beta(U_{l}) \}$
there exists at least one $q \in \{ 1,...,\beta(U_{l+1}) \}$ such that
$\bar{Q}_{p}(U_{l})\cap \bar{Q}_{q}(U_{l+1}) \neq 0$. Therefore we can find a sequence
$i_{2},...,i_{k}$ such that $(i_{1},U_{1}) \sim (i_{2},U_{2}) \sim \cdots
\sim (i_{k},U_{k})=(i_{k},U)$. Hence $i_{k} \in \{i:(i,U) \in I_{a}\}$. \\
{\bf q.e.d.} \\

By Lemma \ref{lemma:partition}, we may define a non-trivial distribution $Q_{a}(U)$  on $U$ by
\[ Q_{a}(U) :=\bigoplus_{\{i:(i,U) \in I_{a}\}} \bar{Q}_{i}(U)
               = \bigoplus_{\{i:(i,U) \in I_{a}\}} span \{Y_{j}: j\in P_{i}(U)\}\] 
for all $a\in \{1,...,A\}$ and $U\in {\cal U}$.  
Define the distributions $Q_{a}$ on $M$ by specifying their restriction  on each 
$U\in {\cal U}$ to be
\[ Q_{a}|_{U} := Q_{a}(U)\] 

\begin{lemma} \label{lemma:welldefined}
The $Q_{a}$ are well-defined. 
\end{lemma}
{\bf Proof:}\\
Let $V$ be another set in ${\cal U}$ having non-zero intersection with $U$. 
We must show that on the intersection, $Q_{a}(U)|_{U\cap V} = Q_{a}(V)|_{U\cap V}$.
By Lemma \ref{lemma:subset}, the set $\{1,..., \beta(U\cap V)\}$
partitions into $\Gamma_{U}(1),...,\Gamma_{U}(\beta(U))$ in such a way that 
\[ \bar{Q}_{i}(U)|_{U\cap V}=\bigoplus_{q\in \Gamma_{U}(i)}\bar{Q}_{q}(U\cap V) \] 
$\{1,..., \beta(U\cap V)\}$ also partitions into $\Gamma_{V}(1),...,\Gamma_{V}(\beta(V))$ 
in such a way that 
\[ \bar{Q}_{j}(V)|_{U\cap V} =\bigoplus_{q\in \Gamma_{V}(j)}\bar{Q}_{q}(U\cap V)\]
Observe that if $\Gamma_{U}(i) \cap \Gamma_{V}(j)\neq \emptyset$ then 
$\bar{Q}_{i}(U)|_{U\cap V} \cap \bar{Q}_{j}(V)|_{U\cap V} \neq 0$ and so
$(i,U) \sim (j,V)$.
Hence,
\begin{eqnarray*}
Q_{a}(U)|_{U\cap V} & := \hspace*{.1in} & 
     \bigoplus_{\{i:(i,U)\in I_{a}\}}\bar{Q}_{i}(U)|_{U\cap V} \\
& =\hspace*{.1in}  & 
\bigoplus_{\{i:(i,U)\in I_{a}\}} \hspace{0.1in} \bigoplus_{q\in \Gamma_{U}(i)} \bar{Q}_{q}(U\cap V) \\
& \subseteq \hspace*{.1in} & 
\sum_{\{i:(i,U)\in I_{a}\}} \hspace{0.1in} 
  \sum_{\{j: \Gamma_{U}(i) \cap \Gamma_{V}(j)\neq \emptyset \}}
\hspace{0.1in} \sum_{q\in \Gamma_{V}(j)} \bar{Q}_{q}(U\cap V)\\
& \subseteq \hspace*{.1in} &  
\sum_{\{i:(i,U)\in I_{a}\}} \hspace{0.1in} \sum_{\{j:(i,U) \sim (j,V)\}}
\hspace{0.1in} \sum_{q\in \Gamma_{V}(j)} \bar{Q}_{q}(U\cap V) \\
& = \hspace*{.1in} & 
\bigoplus_{\{j:(j,V)\in I_{a}\}} \hspace{0.1in} \bigoplus_{q\in \Gamma_{V}(j)} \bar{Q}_{q}(U\cap V)\\
& = \hspace*{.1in} & 
\bigoplus_{\{j:(j,V)\in I_{a}\}} \hspace{0.1in} \bar{Q}_{j}(V)|_{U\cap V}  \\
& := \hspace*{.1in} & 
Q_{a}(V)|_{U\cap V} 
\end{eqnarray*}
Similarly, $Q_{a}(V)|_{U\cap V} \subseteq Q_{a}(U)|_{U\cap V}$ and so the two 
distributions are equal on $U\cap V$. \\
{\bf q.e.d.} \\

The tangent space has the direct sum decomposition into subbundles:
 \[TM=Q_{1}\oplus \cdots \oplus Q_{A} \]
Denote the dual subbundles by $Q_{a}^{*}$ and the fibre of $Q_{a}^{*}$ over $m\in M$ 
by $Q_{a}^{*}(m)$. Then the cotangent space at $m$ has the decomposition:
\[T^{*}_{m}M=Q_{1}^{*}(m)\oplus \cdots \oplus Q_{A}^{*}(m) \]

$h$ is a section of $TM\otimes TM$ and therefore determines a bilinear map
$h(m):T^{*}_{m}M\times T^{*}_{m}M \rightarrow \Re$ for each $m\in M$.
Define positive semi-definite metrics $h_{a}$ on $M$ for $1\leq a\leq A$  by 
\[ h_{a}(m)|_{Q^{*}_{b}(m) \times Q^{*}_{c}(m)} := 
         \delta_{ab}\delta_{ac} h(m)|_{Q^{*}_{a}(m)\times Q^{*}_{a}(m)} \] 
         for each $m\in M$.
On $U \in {\cal U}$, 
\[ h_{a}|_{U}=\sum_{\{ k:(k,U)\in I_{a}\}}h_{k}(U) \]
Therefore $h= \sum_{a=1}^{A}h_{a}.$ 

We may now describe the parallel metrics of $\nabla$ on $M$.

\begin{theorem} \label{theorem:main}
$g$ is a parallel metric on $M$ if and only if it can be written as
$g= \sum_{a=1}^{A}c_{a}h_{a}$ for some $c_{a} \in \Re^{+}$. 
\end{theorem}
{\bf Proof:}\\
$\Longleftarrow$ By Lemma \ref{lemma:a} and the fact that 
$h_{a}|_{U}=\sum_{\{ k:(k,U)\in I_{a}\}}h_{k}(U)$, $\nabla h_{a} = 0$ 
for all $1\leq a \leq A$. Therefore any 
$g= \sum_{a=1}^{A}c_{a}h_{a}$, where $c_{a} \in \Re^{+}$,
is a parallel metric on $M$. \\

\hspace{-0.28in} 
$\Longrightarrow$ Suppose that $g$ is a parallel metric on $M$. 
Let $U$ and $V$ be two sets in ${\cal U}$ with non-empty intersection.
By Lemma
\ref{lemma:subset}, the set $\{1,..., \beta(U\cap V)\}$
partitions into $\Gamma_{U}(1),...,\Gamma_{U}(\beta(U))$ in such a way that 
\[ \bar{Q}_{i}(U)|_{U\cap V}=\bigoplus_{q\in \Gamma_{U}(i)}\bar{Q}_{q}(U\cap V) \]
and 
\[ h_{i}(U)|_{U\cap V}=\sum_{q\in \Gamma_{U}(i)} h_{q}(U\cap V) \] 
$\{1,..., \beta(U\cap V)\}$ also partitions into $\Gamma_{V}(1),...,\Gamma_{V}(\beta(V))$ 
in such a way that 
\[ \bar{Q}_{j}(V)|_{U\cap V} =\bigoplus_{q\in \Gamma_{V}(j)}\bar{Q}_{q}(U\cap V)\]
and
\[ h_{j}(V)|_{U\cap V} =\sum_{q\in \Gamma_{V}(j)} h_{q}(U\cap V)\]
Suppose that $\bar{Q}_{i}(U)\cap \bar{Q}_{j}(V) \neq 0$ for some $i$ and $j$. 
Then there exists an element $p\in \Gamma_{U}(i)\cap \Gamma_{V}(j).$

For any $W\in {\cal U}$, Lemma \ref{lemma:linear} defines positive constants
$c_{k}(W)$ by 
\[ g_{|_{W}}= \sum_{k=1}^{\beta(W)}c_{k}(W)h_{k}(W)\]
Therefore on $U\cap V$, $g$ may be expressed as
\[  g|_{U\cap V}  =  \sum_{k=1}^{\beta(U)}c_{k}(U)h_{k}(U)|_{U\cap V}
     =  \sum_{k=1}^{\beta(U)}c_{k}(U)\sum_{q \in \Gamma_{U}(k)}h_{q}(U\cap V)  \]
Similarly, 
\[  g|_{U\cap V}  =  \sum_{l=1}^{\beta(V)}c_{l}(V)h_{l}(V)|_{U\cap V}
     =  \sum_{l=1}^{\beta(V)}c_{l}(V)\sum_{q \in \Gamma_{V}(l)}h_{q}(U\cap V)  \]
The coefficient of $h_{p}(U\cap V)$ is $c_{i}(U)=c_{j}(V)$. Therefore
\[ \bar{Q}_{i}(U)\cap \bar{Q}_{j}(V) \neq 0 \Longrightarrow c_{i}(U)=c_{j}(V) \]
From the definition of the equivalence $\sim$ it follows that 
\[ (i,U_{1}) \sim (j,U_{2}) \Longrightarrow c_{i}(U_{1})=c_{j}(U_{2}) \]
This allows us to define the positive constants $c_{a} := c_{k}(U)$, where 
$(k,U)$ is any representative of $I_{a}$, for $1\leq a \leq A$.

Thus,
\begin{eqnarray*}
g_{|_{U}}& = \hspace{0.1in} &  \sum_{k=1}^{\beta(U)}c_{k}(U)h_{k}(U) \\
& = \hspace{0.1in} & \sum_{a=1}^{A} \hspace{0.1in}  \sum_{\{ k:(k,U) \in I_{a}\}} 
    c_{k}(U)h_{k}(U)\\
& = \hspace{0.1in} & \sum_{a=1}^{A} \hspace{0.1in} c_{a} \sum_{\{k:(k,U) \in I_{a}\}} 
    h_{k}(U) \\
& = \hspace{0.1in} &  \sum_{a=1}^{A}c_{a}h_{a}|_{U} 
\end{eqnarray*}
It follows that $g= \sum_{a=1}^{A}c_{a}h_{a}$. \\
{\bf q.e.d.}

\begin{corollary} \label{corollary:dimension}
The manifold of parallel metrics on $M$ has dimension $A$.
\end{corollary}

Given the Riemannian manifold $(M,h)$ with generic Levi-Civita connection $\nabla$, 
the determination of the $h_{a}$
is an algebraic construction. Therefore Theorem \ref{theorem:main} enables one to  obtain
all parallel metrics of $\nabla$ by purely algebraic means; integration of differential
equations is unnecessary. \\

Let us summarize the steps in the procedure.\\
1. Find an open cover ${\cal U} \subseteq \widehat{\cal U}$ of $M$  
having the property that the intersection of pairs of sets in ${\cal U}$ is connected. 

For each $U\in {\cal U}$, follow steps 2-5:\\
2. Find a frame of eigenvector fields $(Z_{1},...,Z_{n})$ on $U$ for $R(\xi_{1},\xi_{2})$.\\
3. Obtain the associated frame of orthogonal vector fields $(X_{1},...,X_{n})$. \\
4. Define the orthonormal frame $(Y_{1},...,Y_{n})$. \\
5. Obtain the decomposition $TU=\bar{Q}_{1}(U)\oplus \cdots \oplus \bar{Q}_{\beta(U)}(U)$.\\
6. Construct the subbundles $Q_{a}$. \\
7. Define the tensor fields $h_{a}$ and apply the theorem. \\

\hspace{-0.3in} {\bf Example} Consider the Riemannian manifold ($M,h)$ where
$M:= \Re^{4}$ and $h:= dx^{2}+e^{2x}dy^{2}+du^{2}+e^{2u}dv^{2}$, which in contravariant
form is 
$\frac{\partial}{\partial x}\otimes \frac{\partial}{\partial x}
+ e^{-2x}\frac{\partial}{\partial y}\otimes \frac{\partial}{\partial y}+
\frac{\partial}{\partial u}\otimes \frac{\partial}{\partial u}
+ e^{-2u}\frac{\partial}{\partial v}\otimes \frac{\partial}{\partial v}$.
$(M,h)$ is a Cartesian product of two isomorphic irreducible Riemannian manifolds and therefore
it is expected that the general parallel metric would be a positive-definite 
linear combination of the pullbacks onto $M$ of the component metrics,  
$\frac{\partial}{\partial x}\otimes \frac{\partial}{\partial x}
+ e^{-2x}\frac{\partial}{\partial y}\otimes \frac{\partial}{\partial y}$ and 
$\frac{\partial}{\partial u}\otimes \frac{\partial}{\partial u}
+ e^{-2u}\frac{\partial}{\partial v}\otimes \frac{\partial}{\partial v}$.

The Christoffel symbols for the Levi-Civita connection $\nabla$ of $h$ are
\[ \begin{array}{llc}
\Gamma^{x}_{yy}  & = &  -e^{2x}\\
\Gamma^{y}_{xy} = \Gamma^{y}_{yx} & = & 1 \\
\Gamma^{u}_{vv}  & = &  -e^{2u}\\
\Gamma^{v}_{uv} = \Gamma^{v}_{vu} & =  & 1 
\end{array} \]
and all others zero. Define a good cover ${\cal U}:= \{U,U' \}$ of $M$ by
\[ \begin{array}{lll}
 U & := & \{ (x,y,u,v) \in \Re^{4} : x > u \} \\
 U' & := & \{ (x,y,u,v) \in \Re^{4} : x < u + log 2\} 
   \end{array} \]

On $U$ let 
$ \xi_{1} := \frac{\partial}{\partial x} + \frac{\partial}{\partial u}$
and $\xi_{2} := \frac{\partial}{\partial y} + \frac{\partial}{\partial v}.$
The Riemann curvature with respect to the frame 
$(\frac{\partial}{\partial x},  \frac{\partial}{\partial y},
\frac{\partial}{\partial u}, \frac{\partial}{\partial v}) $
on $U$ is
\[ R(\xi_{1},\xi_{2}) = \left(
\begin{array}{cc|cc}
0 & -e^{2x} & 0 & 0 \\
1 & 0 & 0 & 0 \\ \hline
0 & 0 & 0 & -e^{2u} \\
0 & 0 & 1 & 0 
\end{array} \right) \]
The eigenvalues are $\lambda = -ie^{x}, ie^{x}, -ie^{u}$ and $ie^{u}$,
which are distinct on $U$.
Corresponding eigenvector fields on $U$ are
\begin{center} 
   $Z_{1}= e^{x}\frac{\partial}{\partial x}+i\frac{\partial}{\partial y}, \hspace{0.1in}
   Z_{2}= e^{x}\frac{\partial}{\partial x}-i\frac{\partial}{\partial y}, \hspace{0.1in}
   Z_{3}= e^{u}\frac{\partial}{\partial u}+i\frac{\partial}{\partial v}, \hspace{0.1in}
   Z_{4}= e^{u}\frac{\partial}{\partial u}-i\frac{\partial}{\partial v}$ 
\end{center}   
which defines
\[ \begin{array}{llr}
X_{1} & := & e^{x}  \frac{\partial}{\partial x}|_{U} \\
X_{2} & := &  \frac{\partial}{\partial y}|_{U} \\
X_{3} & := & e^{u}  \frac{\partial}{\partial u}|_{U} \\
X_{4} & := &  \frac{\partial}{\partial v}|_{U}  \end{array} \] 
The orthonormal frame $(Y_{1},...,Y_{4})$ is then given by
\[ \begin{array}{llr}
Y_{1} & := &  \frac{\partial}{\partial x}|_{U} \\
Y_{2} & := &  e^{-x} \frac{\partial}{\partial y}|_{U} \\
Y_{3} & := &  \frac{\partial}{\partial u}|_{U} \\
Y_{4} & := &  e^{-u} \frac{\partial}{\partial v}|_{U}  \end{array} \] 
The curvature form $\omega = \omega^{i}_{j}$ with respect to $(Y_{1},...,Y_{4})$ satisfies
$\omega_{1}^{2}(Y_{2}) = \omega_{3}^{4}(Y_{4}) = 1$. Therefore
$2r(U)1$ and $4r(U)3$. Furthermore, $\omega^{i}_{j}= 0$ for $i\in \{1,2\}$
and $j\in \{3,4\}$. Thus there are exactly two equivalence classes
for the equivalence relation $r(U)$: 
\[ P_{1}(U) = \{1,2 \} \hspace{0.5in} and \hspace{0.5in} P_{2}(U) = \{3,4 \} \] 
This gives,
\[ \begin{array}{lllll}
\bar{Q}_{1}(U) & = & span \{Y_{1}, Y_{2} \} & = & span \{ 
               \frac{\partial}{\partial x}, \frac{\partial}{\partial y} \} |_{U} \\
\bar{Q}_{2}(U) & = & span \{Y_{3}, Y_{4} \} & = & span \{ 
               \frac{\partial}{\partial u}, \frac{\partial}{\partial v} \} |_{U} \\
h_{1}(U) & = & Y_{1}\otimes Y_{1} + Y_{2} \otimes Y_{2} & = & 
  (\frac{\partial}{\partial x}\otimes \frac{\partial}{\partial x}
+ e^{-2x}\frac{\partial}{\partial y}\otimes \frac{\partial}{\partial y})|_{U}\\
h_{2}(U) & = & Y_{3}\otimes Y_{3} + Y_{4} \otimes Y_{4} & = & 
  (\frac{\partial}{\partial u}\otimes \frac{\partial}{\partial u}
+ e^{-2u}\frac{\partial}{\partial v}\otimes \frac{\partial}{\partial v})|_{U}              
    \end{array} \]

On $U'$ let 
$ \xi'_{1} := \frac{\partial}{\partial x} + 2\frac{\partial}{\partial u}$
and $\xi'_{2} := \frac{\partial}{\partial y} + \frac{\partial}{\partial v}.$
The Riemann curvature with respect to the frame 
$(\frac{\partial}{\partial x},  \frac{\partial}{\partial y},
\frac{\partial}{\partial u}, \frac{\partial}{\partial v}) $
on $U'$ is
\[ R(\xi'_{1},\xi'_{2}) = \left(
\begin{array}{cc|cc}
0 & -e^{2x} & 0 & 0 \\
1 & 0 & 0 & 0 \\ \hline
0 & 0 & 0 & -2e^{2u} \\
0 & 0 & 2 & 0 
\end{array} \right) \]
The eigenvalues are $\lambda' = -ie^{x}, ie^{x}, -2ie^{u}$ and $2ie^{u}$, 
which are distinct on $U'$.
Corresponding eigenvector fields on $U'$ are
\begin{center} 
$Z'_{1}= e^{x}\frac{\partial}{\partial x}+i\frac{\partial}{\partial y}, \hspace{0.1in}
Z'_{2}= e^{x}\frac{\partial}{\partial x}-i\frac{\partial}{\partial y}, \hspace{0.1in}
Z'_{3}= e^{u}\frac{\partial}{\partial u}+i\frac{\partial}{\partial v}, \hspace{0.1in}
Z'_{4}= e^{u}\frac{\partial}{\partial u}-i\frac{\partial}{\partial v}$ 
\end{center}   
which defines
\[ \begin{array}{llr}
X'_{1} & := & e^{x}  \frac{\partial}{\partial x}|_{U'} \\
X'_{2} & := &  \frac{\partial}{\partial y}|_{U'} \\
X'_{3} & := & e^{u}  \frac{\partial}{\partial u}|_{U'} \\
X'_{4} & := &  \frac{\partial}{\partial v}|_{U'}  \end{array} \] 
The orthonormal frame $(Y'_{1},...,Y'_{4})$ is then given by
\[ \begin{array}{llr}
Y'_{1} & := &  \frac{\partial}{\partial x}|_{U'} \\
Y'_{2} & := &  e^{-x} \frac{\partial}{\partial y}|_{U'} \\
Y'_{3} & := &  \frac{\partial}{\partial u}|_{U'} \\
Y'_{4} & := &  e^{-u} \frac{\partial}{\partial v}|_{U'}  \end{array} \] 
Continuing the analysis as above gives
\[ P_{1}(U') = \{1,2 \} \hspace{0.5in} and \hspace{0.5in} P_{2}(U') = \{3,4 \} \] 
and
\[ \begin{array}{lll}
\bar{Q}_{1}(U') & = & span \{ 
               \frac{\partial}{\partial x}, \frac{\partial}{\partial y} \} |_{U'} \\
\bar{Q}_{2}(U') & = & span \{ 
               \frac{\partial}{\partial u}, \frac{\partial}{\partial v} \} |_{U'} \\
h_{1}(U') & = &   
(\frac{\partial}{\partial x}\otimes \frac{\partial}{\partial x}
+ e^{-2x}\frac{\partial}{\partial y}\otimes \frac{\partial}{\partial y})|_{U'}\\
h_{2}(U') & = &  
(\frac{\partial}{\partial u}\otimes \frac{\partial}{\partial u}
+ e^{-2u}\frac{\partial}{\partial v}\otimes \frac{\partial}{\partial v})|_{U'} 
   \end{array} \]

Next we consider the equivalence relation $\sim$. Restricted to $U\cap U'$,
\[ \begin{array}{lll}
\bar{Q}_{1}(U)|_{U\cap U'} & = &  \bar{Q}_{1}(U')|_{U\cap U'} \\
\bar{Q}_{2}(U)|_{U\cap U'} & = &  \bar{Q}_{2}(U')|_{U\cap U'} \\
\bar{Q}_{1}(U) \cap \bar{Q}_{2}(U') & = & 0 \\
\bar{Q}_{2}(U) \cap \bar{Q}_{1}(U') & = & 0
   \end{array} \]
Therefore $(1,U) \sim (1,U')$, $(2,U) \sim (2,U')$ and there are no other non-trivial
relations. This gives
\[ \begin{array}{lll}
  I_{1} & = & \{ (1,U), (1,U') \} \\
  I_{2} & = & \{ (2,U), (2,U') \} \\
  Q_{1} & = & span \{  \frac{\partial}{\partial x}, \frac{\partial}{\partial y} \} \\
  Q_{2} & = & span \{  \frac{\partial}{\partial u}, \frac{\partial}{\partial v} \} \\
  h_{1} & = & \frac{\partial}{\partial x}\otimes \frac{\partial}{\partial x}
+ e^{-2x}\frac{\partial}{\partial y}\otimes \frac{\partial}{\partial y} \\
  h_{2} & = & \frac{\partial}{\partial u}\otimes \frac{\partial}{\partial u}
+ e^{-2u}\frac{\partial}{\partial v}\otimes \frac{\partial}{\partial v}
  \end{array} \]
By Theorem \ref{theorem:main}, $g$ is a parallel metric with respect to $\nabla$
if and only if $g=c_{1}h_{1}+c_{2}h_{2}$ for some constants $c_{1}, c_{2} \in \Re^{+}$;
the anticipated result.

\end{document}